\documentclass[useAMS,usenatbib]{mn2e}

\usepackage{dsfont}
\usepackage{amsmath}
\usepackage{amssymb}
\usepackage{graphicx}
\usepackage{verbatim}
\usepackage{natbib}
\usepackage{eucal}
\usepackage{calligra}

\usepackage{amsfonts}
\usepackage[usenames,dvipsnames]{color}
\usepackage[normalem]{ulem}
\usepackage[T1]{fontenc}

\DeclareMathAlphabet{\mathscr}{OT1}{pzc}%
                                 {m}{it}

\newcommand{\mnras}{MNRAS}
\newcommand{\jcap}{JCAP}
\newcommand{\apj}{ApJ}

\newcommand{\nat}{Nature (London)}

\def\apjl{ApJL}

\newcommand{\be}{\begin{equation}}
\newcommand{\ee}{\end{equation}}
\newcommand{\bes}{\begin{equation*}}
\newcommand{\ees}{\end{equation*}}
\newcommand{\bea}{\begin{eqnarray}}
\newcommand{\eea}{\end{eqnarray}}
\newcommand{\beas}{\begin{eqnarray*}}
\newcommand{\eeas}{\end{eqnarray*}}

\newcommand{\mpch}{\;{\rm Mpc}/h}

\newcommand{\reff}{r_{\rm eff}}
\newcommand{\xivv}{\xi_{\rm vv}}
\newcommand{\xivg}{\xi_{\rm vg}}
\newcommand{\xigg}{\xi_{\rm gg}}
\newcommand{\ximm}{\xi_{\rm mm}}
\newcommand{\bvrm}{b_{\rm v}}
\newcommand{\bvauto}{b_{\rm v}^{\rm auto}}
\newcommand{\bvcross}{b_{\rm v}^{\rm cross}}
\newcommand{\bg}{b_{\rm g}}

\begin{document}

\title[Void Clustering in SDSS]
	{Clustering and Bias Measurements of SDSS Voids}
\author[Clampitt, Jain and S\'{a}nchez]
	{Joseph~Clampitt$^{1,}$\thanks{E-Mail: clampitt@sas.upenn.edu}, Bhuvnesh~Jain$^{1}$, Carles~S\'{a}nchez$^{2}$ \\
	$^1$Department of Physics and Astronomy, Center for Particle Cosmology, \\
  	University of Pennsylvania,
  	209 S. 33rd St., Philadelphia, PA 19104, USA \\
	$^2$Institut de F\'{i}sica d'Altes Energies, Universitat Aut\`{o}noma de Barcelona, E-08193 Bellaterra (Barcelona), Spain}
	
\maketitle

\begin{abstract}
Using a void catalog from the SDSS survey, we present the first measurements of void clustering and the corresponding void bias. Over the range $30-200 \mpch$ the void auto-correlation is detected at $5\sigma$ significance for voids of radius $15-20 \mpch$.
We also measure the void-galaxy cross-correlation at  higher signal-to-noise and compare the inferred 
void bias with the autocorrelation results.
Void bias is constant with scale for voids of a given size, but  its value falls from $5.6 \pm 1.0$ to below zero as the void radius increases from $15$ to $30\mpch$.
The comparison of our measurements with carefully matched galaxy mock catalogs, with no free parameters related to the voids, shows that model predictions can be reliably made for void correlations. We study the dependence of void bias on tracer density and void size with a view to future applications. 
In combination with our previous lensing measurements of void mass profiles, these clustering measurements provide another step towards using voids as cosmological tracers.
\end{abstract}

\begin{keywords}
cosmology: observations -- dark matter -- large scale structure of Universe
\end{keywords}

\section{Introduction}

The clustering of galaxies was first detected over 60 years ago \citep{l54,r54}, and since then this observable has developed into one of the cornerstones of modern cosmology.
While the abundance and other properties of voids have been measured in the past, 
 this paper is the first successful attempt to detect their clustering in survey data.
Galaxy clustering is strongest on nonlinear scales less than 10 $\mpch$, but these scales are smaller than the radius of a typical void, so that meaningful void-void correlations are impossible to measure at such scales.
Only the large scale regime where the magnitude of 2-point correlations has fallen to the $\sim 0.1$ level and below is available for void clustering measurements.
Not only is the available signal smaller, but the noise in void correlation functions is greater since voids are far less numerous than galaxies. 
Voids are much larger and finding even one requires many galaxies to trace its shape.

Theoretical work on modeling the bias of dark matter halos \citep{ck89,mw96,st99} was extended to voids by \citet{sv04}.
The resulting bias function starts out positive for the smallest voids but eventually transitions to negative values for the largest objects.
\citet{hws14} measured the void-void and void-galaxy power spectrum in simulations, recovering this transition to negative bias for larger voids.
(See also \citet{pcl05} for an earlier study of void-void correlations in simulations.)
Using voids traced by simulated dark matter particles, \citet{chd14} showed that this transition scale also depends on void tracer density.
On the observational side, the focus has been on the size distribution of voids and the distribution and properties of galaxies within the voids, not on large scale clustering and void bias.
One exception is the void-galaxy cross-correlation measurements of \citet{plc13}, whose main aim was to understand the dynamics of voids.

In \citet{cj15}, henceforth CJ15, we presented a new void finder and void catalog using Sloan Digital Sky Survey\footnote{http://www.sdss.org} (SDSS) data.
In addition, we measured the density profiles of these voids using weak lensing, finding an average central density $\delta \lesssim -0.5$.
Complementing previous void profiles assumed from visible galaxies \citep{hv04, pvh12, cpl13, slh14, nhd15}, our lensing measurements showed directly that these objects are voids in the dark matter.
In this work, we use the same set of voids to measure void clustering and bias.
In Sec 2. we describe our SDSS and mock data sets.
In Sec. 3 we present high-significance measurements of void-void clustering, void-galaxy clustering, and void bias, and make comparison with mock measurements of the same.
Finally, in Sec. 4 we discuss the implications of the measurements and possible future directions for void clustering studies.

\section{SDSS and Mock Data Sets}

\begin{table*}
\centering
Void Tracer Properties of SDSS LRGs and Simulation Halos\\
\begin{tabular}{l|c|c|c|c|c}
\hline
sample name & dataset & density & inter-particle spacing & halo mass threshold & halo bias \\
\hline
Data & SDSS & $1\times10^{-4} \, ({\rm Mpc}/h)^{-3}$ & 21.5 $\mpch$ & - & 2.1 \\
Matched & HR simulation & $1\times10^{-4} \, ({\rm Mpc}/h)^{-3}$ & 21.5 $\mpch$ & $1.3\times 10^{13} M_\odot / h$ & 2.1 \\
Sparse & HR simulation & $1\times10^{-4} \, ({\rm Mpc}/h)^{-3}$ & 21.5 $\mpch$ & $2.1\times 10^{13} M_\odot / h$ & 1.8 \\
Medium & HR simulation & $2\times10^{-4} \, ({\rm Mpc}/h)^{-3}$ & 17.1 $\mpch$ & $1.3\times 10^{13} M_\odot / h$ & 2.1 \\
Dense & HR simulation & $4\times10^{-4} \, ({\rm Mpc}/h)^{-3}$ & 13.5 $\mpch$ & $0.7\times 10^{13} M_\odot / h$ & 2.4
\end{tabular}
\caption{Description of the SDSS galaxies and HR simulated halos used to define voids.
The halo mass threshold of the mock samples indicates the minimum halo mass included.
The Matched sample of simulated halos is our primary sample for comparison to the Data.
}
\end{table*}

\subsection{SDSS void catalog}

For void tracers in SDSS we use the Luminous Red Galaxy (LRG) sample of \citet{kbs10}, 
specifically the volume limited sample at $0.16 < z < 0.36$.
We use the largest contiguous patch ($\sim$ 7,000 square degrees) of the available data.
This sample has $\sim$ 56,000 LRGs and covers a comoving volume $\sim 0.6 \, ({\rm Gpc}/h)^3$.
See Table 1 for more details on this sample.

The void finder of CJ15 begins by dividing the SDSS LRG sample into redshift slices 
of various thickness. Within each of these slices, all LRGs are projected onto a 
2D spherical space and holes in the LRG distribution are found within each slice.
These potential void centers are then culled with several quality cuts on the void axis ratio,
location (required to be well within the survey edges), and volume overlap between voids.
CJ15 showed that the density profiles derived from weak lensing were consistent between samples with volume overlap cutoffs of 50\% and 90\%.
For most of our results, we require that each void has an overlapping volume fraction of 50\% or less; this simplifies somewhat the interpretation of the clustering measurements as coming from unique voids.
This fiducial sample has $\sim$11,000 voids.
In Sec.~\ref{sec:overlap} we briefly compare the void-galaxy correlation using the 90\% overlap sample, with 19,000 objects this sample increases the number of voids by about a factor of 2.

The void finder outputs two sizes for each void: a projected radius in the plane of the sky ($R_{\rm v}$) and 
a size in the line-of-sight direction ($s_{\rm v}$).
We expect voids with the same clustering amplitude and bias to have comparable total volume.
Thus we define an effective radius which is the radius of the sphere with the same 
volume as the ellipsoid output by the void finder:
\be
\reff \equiv (R_{\rm v}^2 s_{\rm v})^{1/3} \, .
\ee
For all measurements, we group the voids in three size bins: $\reff = 15-20\mpch$, $20-25\mpch$, and $25-30\mpch$.

\subsection{Horizon Run mocks}

For mock void samples, we use one of the eight public full-sky mock surveys from the Horizon Run Simulation\footnote{http://sdss.kias.re.kr/astro/Horizon-Run/} \citep{kpg09} with cosmology $\Omega_m = 0.26$, $\Omega_{\Lambda} = 0.74$, and $\sigma_8 = 0.79$.
Restricting to the same redshift range as the data, this gives us $\sim 6$ times the volume.
We make use of four different simulated halo samples as void tracers.
First, the Matched sample is chosen to have the same density and halo bias as the LRGs in the data.
The other three halo samples are chosen by setting minimum mass thresholds such that, in comparison to the data, they have: exactly the same density (Sparse), $2\times$ the density (Medium), and $4\times$ the density (Dense).
The result is that the tracer bias varies by $\sim 15-30\%$ along with the larger variation in tracer density.
We summarize the tracer densities, inter-particle spacing, halo mass thresholds, and halo bias of these samples in Table 1.

Note that our Sparse and Matched sample voids have an average inter-particle spacing of $21.5\mpch$, which is larger than our smallest measured bin of void sizes, $15\mpch < \reff < 20\mpch$.
Although this bin may therefore have some contamination from chance gaps between galaxies (``Poisson'' voids), the lensing measurements of CJ15 indicate that voids in this sample are on average significantly underdense in the dark matter.
Furthermore, for these smaller voids we see a smooth change in void bias (see Sec.~\ref{sec:bias} and Fig.~\ref{fig:bias}) as tracer density is increased from the Sparse to Medium, and finally to the Dense sample (for which all voids are larger than the inter-particle spacing).
This provides further evidence that we do not see a qualitative change from true voids to spurious voids as the tracer inter-particle spacing is increased to be slightly larger than the void size.

\begin{figure*}
\centering
\resizebox{180mm}{!}{\includegraphics{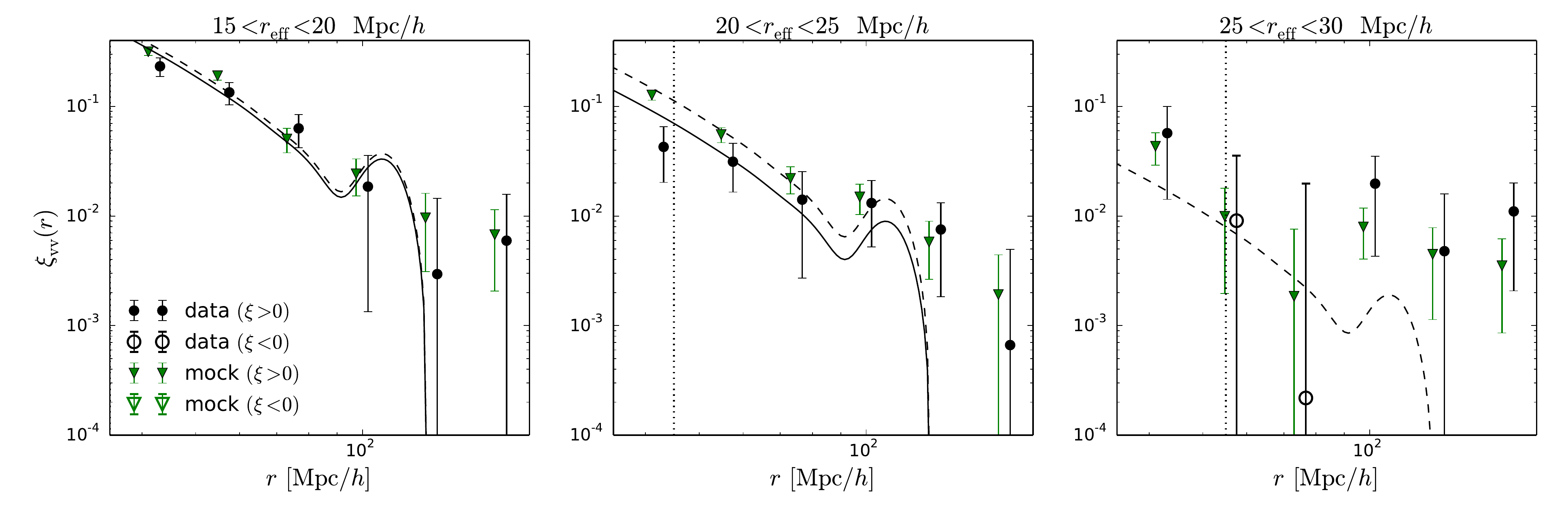}}
\caption{Void-void clustering $\xivv$ using both Data (black circles) and Matched mock catalogs (green triangles).
The data and mocks give qualitatively consistent results outside twice the void radius $2\, \reff$ (vertical dotted line).
There is a visible decrease in clustering amplitude as void radius increases (left to right panels).
As in galaxy-galaxy clustering, $\xivv$ can be written as the 
product of the matter-matter correlation function with the square of the void bias. 
The predicted $\xivv$ with the best-fit values of the linear bias (see Fig.~\ref{fig:bias}) from data (solid line) and mocks (dashed) are shown.
The consistency with linear bias supports the symmetric relationship between 
galaxies and clusters, which form from peaks in the primordial matter field, and voids, formed from troughs.
}
\label{fig:vv-log}
\end{figure*}

\section{Results}

We measure the 2-point correlation function $\xi(r)$ using the estimator of \citet{ls93}:
\be
\xi(r) = \frac{DD(r) - DR(r) - RD(r) + RR(r)}{RR(r)} \, ,
\ee
where, e.g., for void-void clustering $DD(r)$ denotes the number of void-void pairs, $DR(r)$ and $RD(r)$ the number of void-random pairs, and $RR(r)$ the number of random-random pairs with comoving separation $r$.
We use \verb+TreeCorr+ \footnote{https://github.com/rmjarvis/TreeCorr} \citep{jbj04} to compute all correlation functions.
We obtain the covariance by splitting the survey into 128 patches and using the jackknife method described in \citet{nbg09}.
Likewise for the whole-sky simulation covariance, but in that case we use 192 patches, each 4x larger than the data jackknife patches.
We make use of HEALpix\footnote{http://healpix.sf.net} \citep{ghb05} for defining equal-area jackknife patches.
We checked that our error bars are not sensitive to this difference between the size of mock and data patches by repeating the measurements using a set of data patches with comparable size to the mock patches.
For both voids and galaxies, we use corresponding random catalogs with 20 times as many points.

\subsection{Void-void clustering}
\label{sec:vv}

In Fig.~\ref{fig:vv-log} we show the void-void clustering measurement, $\xivv$, in data and mocks.
Outside $2\,\reff$ all correlations involve distinct voids, and in this regime the data and mocks are in qualitative agreement.
To obtain the significance of detection, we use a likelihood ratio test comparing two hypotheses: the null hypothesis of no clustering and the simulation hypothesis that the data was generated from a model with central values and covariance matching the simulation measurements.
Considering only the data on scales $2\,\reff < r < 200 \mpch$, the detection significance for the $\reff = 15-20\mpch$ bin is slightly larger than $5\sigma$, corresponding to a p-value $ \sim 2 \times 10^{-7}$.
For the $20-25\mpch$ bin we obtain a $3\sigma$ measurement, corresponding to a p-value 0.0014.
The largest radius voids are fewest in number, leading to a measurement consistent with both the null hypothesis and the simulations.

The data is strong enough to discern a decrease in the correlation strength with void size, and the mocks show this trend even more clearly.
In \S~\ref{sec:bias} we comment more on these trends, and quantify the variation further with fits to the void bias $\bvrm$.
In Fig.~\ref{fig:vv-log} we also plot our best fit model of the void-void correlation, $\xivv = \bvrm^2 \ximm$ for both data and mocks.
These theoretical curves match the measurement very well, as expected if voids are biased tracers of the dark matter density field.
In addition to comparing this single-parameter model to the data, it is useful to directly compare the data to the Matched sample of simulated voids.
The reduced chi-square (using radial bins $2\reff < r < 200 \mpch$) is 8/6, 4/5, and 2/5 for the three size bins, all acceptable fits.
However, the more sensitive likelihood ratio test shows tension at the $1$ to $2\sigma$ level for the $15-20\mpch$ and $20-25 \mpch$ bins.
Although this tension is relatively small, it may be pointing to a slight mismatch between our Data and Matched simulation samples (see Sec.~\ref{sec:discussion} for further discussion).

\begin{figure*}
\centering
\resizebox{180mm}{!}{\includegraphics{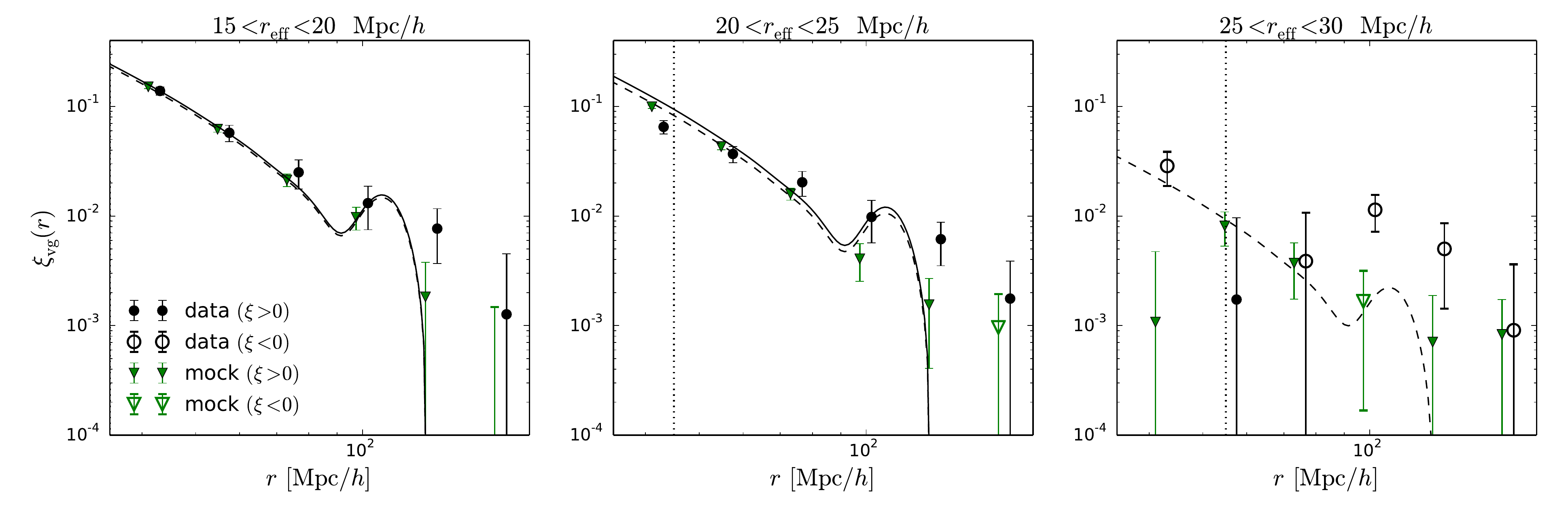}}
\caption{Same as Fig.~\ref{fig:vv-log}, but showing void-galaxy cross-correlation $\xivg$. 
The cross-correlation signal to noise is much higher, since there are $\sim10\times$ more 
galaxies than voids. Nonetheless, data and mock measurements are consistent outside twice the void 
radius (vertical dotted line).
The cross-correlation is expressed as the product of the matter-matter correlation 
function with one factor each of void and galaxy bias (solid and dashed lines).
}
\label{fig:vg-log}
\end{figure*}

\begin{figure*}
\centering
\resizebox{180mm}{!}{\includegraphics{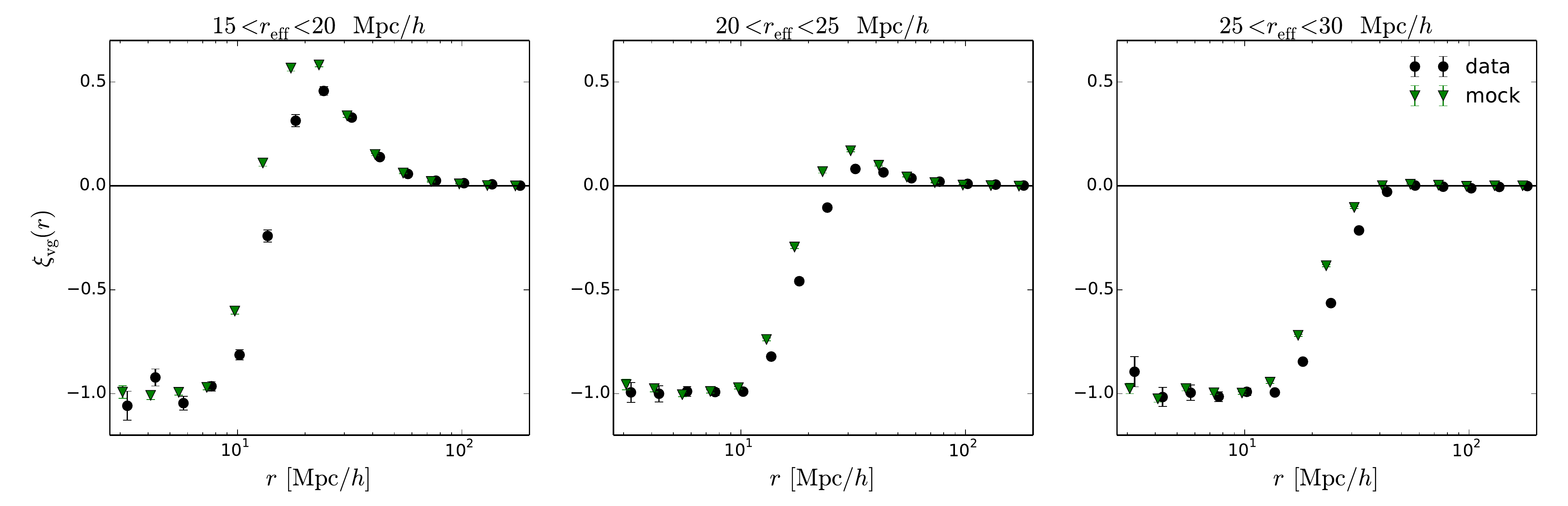}}
\caption{Same as Fig.~\ref{fig:vg-log}, but showing the cross-correlation over a larger range of scales $r$ and with a linear vertical axis.
As expected based on the definition of the void sample, the cross-correlation approaches $-1$ near the void center.
At intermediate scales $\sim \reff/2$ to $2\reff$, the void profile shows a prominent compensatory ridge of galaxies for the smallest voids.
The ridge disappears for the largest voids, which as a result are underdense out to very large scales.}
\label{fig:vg-linear}
\end{figure*}

\subsection{Void-galaxy clustering}

The void-galaxy clustering measurement, $\xivg$, is shown in Fig.~\ref{fig:vg-log}.
Due to the much larger number of galaxies compared to voids ($\sim 10$ times more), the signal to noise (S/N) is much higher in this measurement compared to the auto-correlation.
It thus yields more precise values for void bias (see \S~\ref{sec:bias}).
We plot our best fit model, $\xivg = \bvrm \bg \ximm$ for both data and mocks.
The galaxy bias, defined as $\bg = \sqrt{\xigg / \ximm}$, is $\bg = 2.1$ for the Data LRGs, and $\bg = 2.1$ for our Matched halo sample.
As with $\xivv$, the goodness-of-fit of the simulation model is acceptable for all three void radius bins: in order of increasing void size the reduced $\chi^2$ is 4/6, 6/5, and 7/5.
The tension between mocks and data in $\xivv$ based on the likelihood ratio statistic is not present in the measurement of $\xivg$.
In all three $\reff$ bins, mocks and data are consistent within $\sim 1\sigma$ using the likelihood ratio test (see Sec.~\ref{sec:vv} for details).

In Fig.~\ref{fig:vg-linear} we show the entire range of the void-galaxy clustering measurement: (i) the innermost scales $r < \reff /2$ where $\xivg = -1$ by definition, (ii) the void profile regime between $\reff/2$ and $2 \reff$, and (iii) the linear regime $r > 2 \reff$ (the subject of Fig.~\ref{fig:vg-log}).
Regarding (i), we note simply that both the data and mock measurements are indeed equal to -1 at small scales, a reassuring check of the measurement and random point catalogs.
The void profile in (ii) shows more structure: the smallest voids display a clear positive bump which becomes less prominent as void size $\reff$ increases, and disappears entirely for voids with $\reff > 25 \mpch$.
Thus the largest voids are the only ones that dominate their environments and are underdense out to very large scales.
These results are qualitatively consistent with those of \citet{cpl13} and \citet{hsw14}, who find that larger voids lack the compensatory ridge that surrounds many smaller voids.
(Although see also \citet{nhd15}, on void profiles that show less significant differences as a function of size.)
Due to the sharply rising profile and good signal-to-noise in this regime, even slight differences between data and mocks are easy to see.
Fig.~\ref{fig:vg-linear} shows that for all three void size bins, the mock profiles have a denser and somewhat sharper ridge of galaxies around the void.
In Sec.~\ref{sec:discussion}, we discuss possible differences between data and mocks that could be the cause.
In the following section, we discuss in more detail the linear regime results and implications for void bias.

\subsection{Void bias}
\label{sec:bias}

\begin{figure*}
\centering
\resizebox{85mm}{!}{\includegraphics{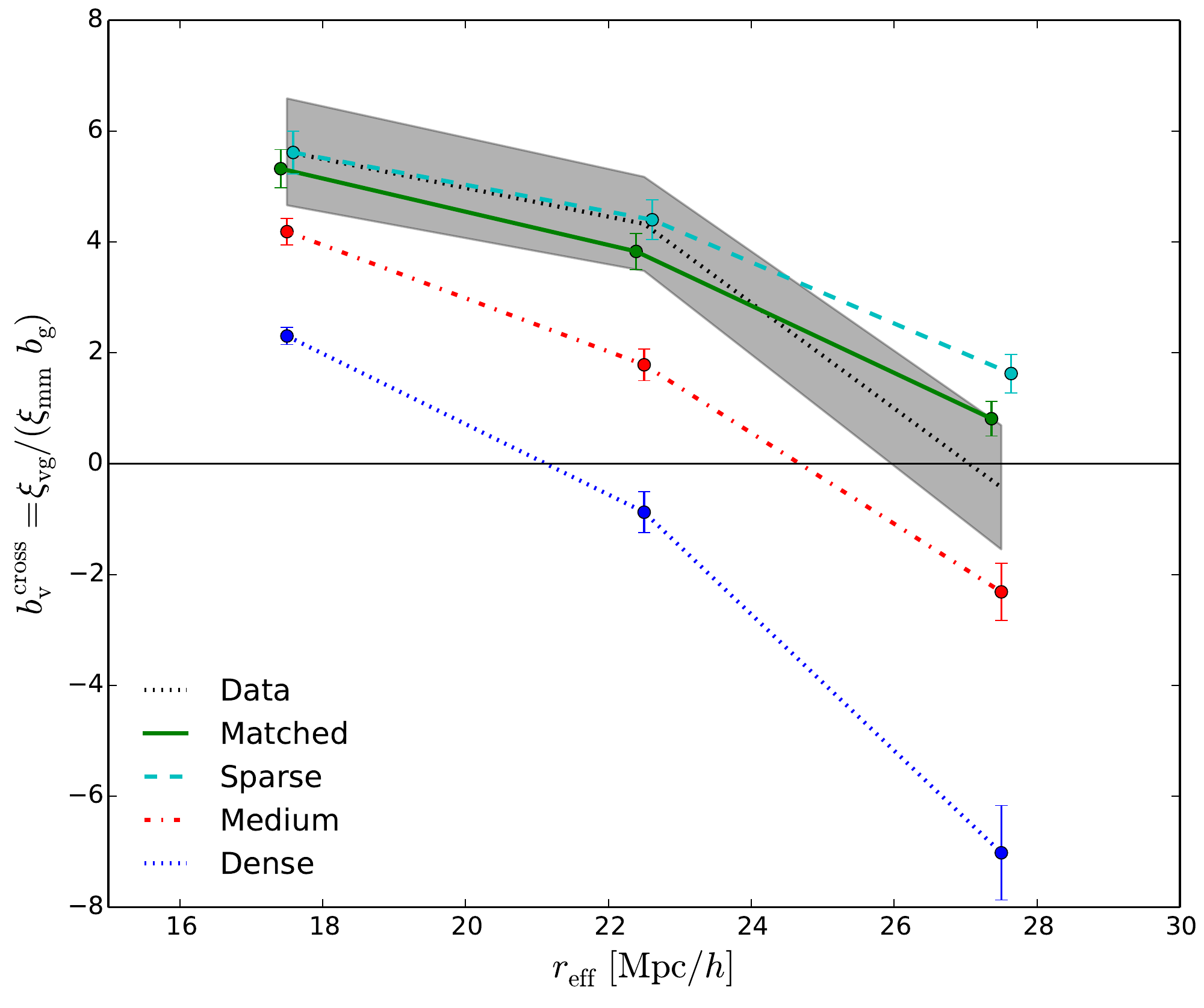}}
\resizebox{85mm}{!}{\includegraphics{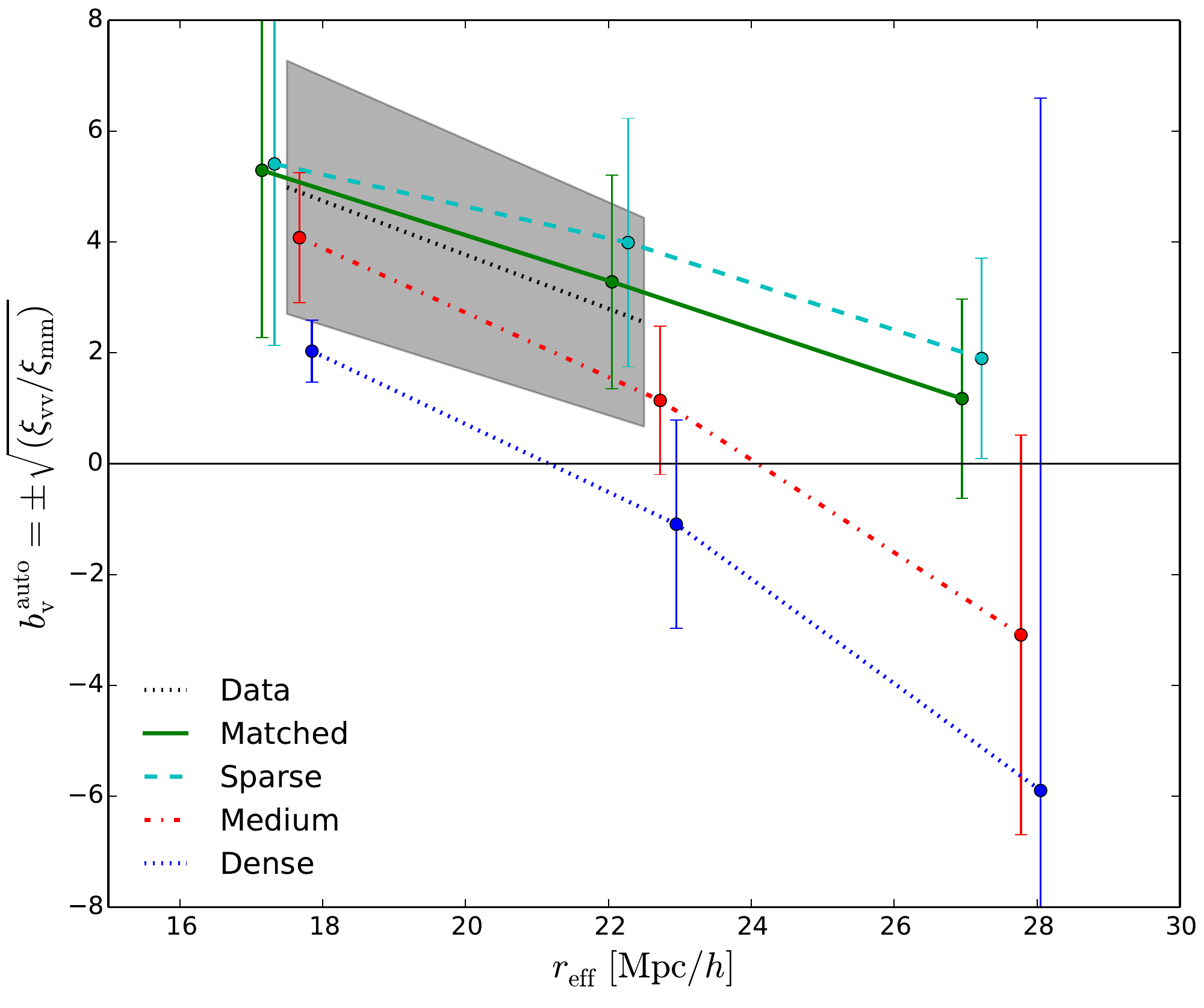}}
\caption{(left panel): Void bias calculated from $\xi_{vg}$, averaged over points between $2\,\reff < r < 80 \mpch$.
The agreement between the Data (gray bands) and Matched mock samples (green circles) is excellent.
The trend towards smaller bias for larger and more densely sampled voids is clear.
Mock samples of various density and bias are as described in the legend.
(right panel): Same as the left panel, but with void bias calculated from $\xi_{vv}$. 
The errors are much larger in this case, but the two different estimators of void bias are consistent.
}
\label{fig:bias}
\end{figure*}

For both data and mocks we obtain two values for the best fit bias of each void size bin:
one via the void-galaxy correlation,
\be
\bvcross = \frac{\xivg}{\bg \ximm} \, ,
\ee
and one via the void-void correlation,
\be
\bvauto = \pm \sqrt{\frac{\xivv}{\ximm}} \, ,
\ee
where $\ximm$ is the matter-matter correlation function measured from the simulations.
With the auto-correlation we actually measure bias squared, then take the square root and choose the sign of bias that matches the sign of $\xivg / (\ximm \bg)$.
Both fits use only the data between $2\,\reff < r < 80 \mpch$: the lower limit assures we use only pairs of distinct voids in the bias measurement, while the upper limit keeps us from dividing by the value of $\ximm$ where it is falling rapidly and approaching the Baryon Acoustic Oscillations (BAO) scale (and thus more strongly dependent on cosmology).
The noise on $\bvauto$ for the largest voids in the data is very large and thus we were unable to obtain a bias fit for that sample.

The resulting bias values are shown in Fig.~\ref{fig:bias}.
The SDSS Data voids clearly show decreasing bias with void radius.
In the left panel, the bias falls from $5.6 \pm 1.0$ to $-0.4 \pm 1.1$ between $\reff = 15-20\mpch$ and $25-30\mpch$, an $\sim 5\sigma$ detection of a decreasing trend with radius.
Comparing between the Data and Matched mock samples, the fitted void bias values are consistent for both $\bvauto$ and $\bvcross$.
Similarly, comparing between the two estimates of bias, both provide reassuringly consistent results in both data and mocks.

The mocks allow us to use higher tracer densities compared to the SDSS LRG sample and explore the effect on void bias.
In all cases, the same galaxies (the Matched sample) are used in the cross-correlation measurement, to isolate the effect of tracer density on void bias.
It is evident that there is a range of bias values, with the maximum positive and negative bias suggesting objects that are as rare as large galaxy clusters.
For the Dense and Medium samples we find an interesting transition of the void bias from positive to negative values with increasing radius.
The void radius at which bias crosses zero depends on tracer density: it occurs at $\sim 21\mpch, \, 25\mpch,$ and above $30\mpch$ for our three samples of decreasing density.
This trend has been studied before in simulations by \citet{hws14} and \citet{chd14}.
Those works used tracer densities of $0.02 \, ({\rm Mpc}/h)^{-3}$ and larger, at least 50 times denser than our samples.
The qualitative dependence of void bias on radius is the same between our voids and those of \citet{hws14} and \citet{chd14}.
However, the detailed values of the bias depend on the void finder, and the fact that our bias zero crossing for the Dense sample is similar to that of \citet{hws14} (at $19.5\mpch$) is a coincidence.

Fixing the void size, \citet{chd14} showed that simulated voids found using denser tracer samples have lower bias.
Here we extend those results to observationally interesting tracer densities and find the same qualitative trends.
Our Matched sample with density $10^{-4} ({\rm Mpc}/h)^{-3}$ has higher bias at all void sizes than the $2\times10^{-4} ({\rm Mpc}/h)^{-3}$ sample, which in turn has higher bias than the $4\times10^{-4} ({\rm Mpc}/h)^{-3}$ sample.
The difference in bias between samples is greatest for the voids with larger radii.
Note that this densest sample is comparable to the density of the CMASS \citep{aab14} and LOWZ \citep{trb14} galaxy samples of the Baryon Oscillation Spectroscopic Survey\footnote{https://www.sdss3.org/surveys/boss.php} (BOSS) \citep{aaa15}.

Finally, we find that the void bias depends on the tracer bias.
The Matched sample has the same tracer density as the Sparse sample, but with an average halo bias that is 15\% smaller.
This smaller halo bias causes the void bias to drop by 5\%, 15\%, and 50\% for the three increasing void radius bins (see Fig.~\ref{fig:bias}).
Thus, even a small change in tracer bias makes a substantial difference in the void bias for larger voids.
The importance of tracer bias on simulated void properties was also recently studied by \citet{pollina16}.
This work points out that due to this dependence on tracer bias, simply subsampling dark matter particles of a simulation to the appropriate density is not sufficient to match observed galaxy-defined void samples.
An explanation for the trends described above is discussed in Sec.~\ref{sec:discussion}.

\begin{figure*}
\centering
\resizebox{180mm}{!}{\includegraphics{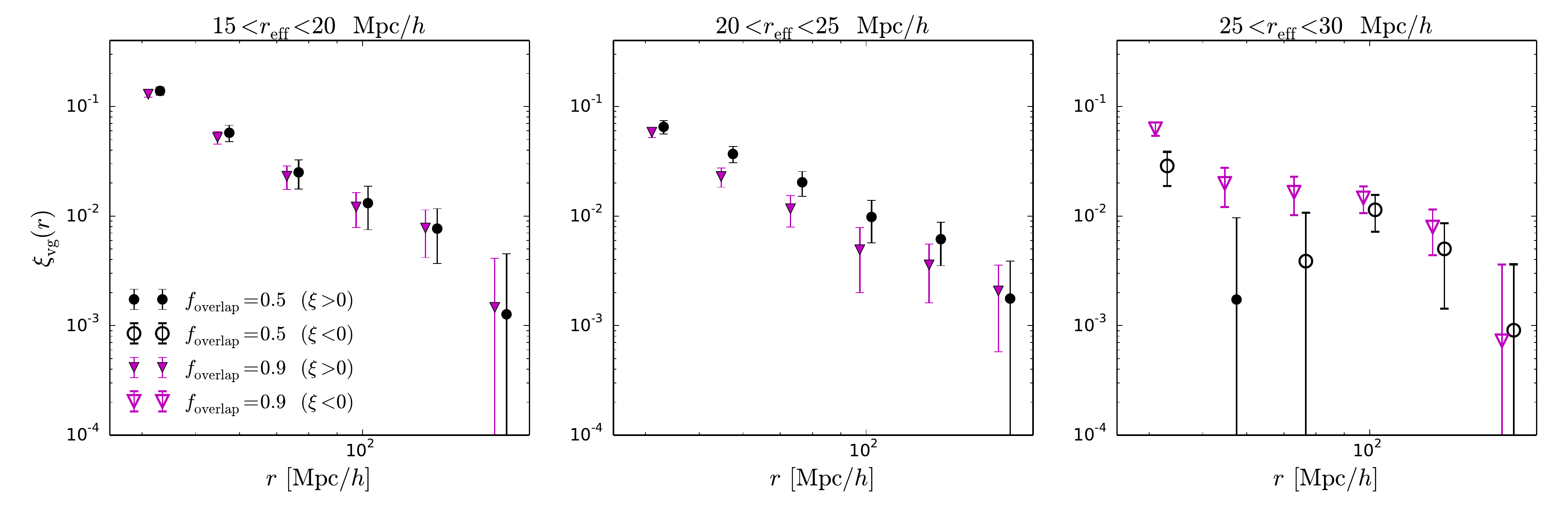}}
\caption{Same as Fig.~\ref{fig:vg-log}, but comparing a sample which allows more overlap between neighboring voids (up to 90\%, magenta triangles) compared to our fiducial sample (up to 50\% overlap, black circles).
Both sets of points are measured in the SDSS data.
Correlations, and therefore void bias, are more negative for the sample which allows more overlap.
The effect on the $\reff = 25-30 \mpch$ sample is especially striking: the measured bias transitions from $\sim 0$ to negative values when the cut on overlapping volume is relaxed.
}
\label{fig:vg-overlap}
\end{figure*}

\subsection{Volume overlap}
\label{sec:overlap}

The void finder of CJ15 allows us to choose the maximum amount of overlap allowed between neighboring voids.
In Fig.~\ref{fig:vg-overlap} we study the effect of this parameter by comparing the measured SDSS void-galaxy correlations using both 90\% maximum overlap between voids and 50\% (our fiducial sample).
The results are consistent for the smallest void sample, but for the medium and largest voids there is a significant difference.
Correlations, and therefore void bias, are more negative for the sample which allows more overlap.
For the $\reff = 20-25 \mpch$ sample the void bias is 30-40\% smaller.
The effect on the $\reff = 25-30 \mpch$ sample is even more striking: the measured bias is clearly negative for the 90\% overlap sample.
This is our first clear detection of a negative void bias in survey data.
While there is a difference in the bias of the two void samples, further study will be needed to verify conclusively whether this is due to the differing void overlap.
The two samples also differ in their statistics: the 90\% overlap sample has a factor of 2 more voids, and hence the error bars are significantly tighter.
We have not ruled out the possibility that the differences above are simply due to better signal-to-noise, and leave further study of volume overlap trends for future work.

There is also some effect in the transition region (not pictured) from $\xivg \sim -1$ to its peak around $r \sim \reff$, with the direction of the effect such that the 90\% overlap sample has a denser and somewhat sharper ridge of galaxies around the void.
Note that the results shown in Fig.~7 of CJ15 for the lensing mass profiles of these two void samples were consistent.
The higher S/N void-galaxy cross-correlation are required in order to pick up this slight difference in the profiles.
Finally, we also compared the two samples' void-void correlations, but again in this case the errors are too large to discern a difference.

\section{Discussion}
\label{sec:discussion}

We have presented a $5\sigma$ detection of the void auto-correlation function (Fig.~\ref{fig:vv-log}) over $30-200 \mpch$ using the SDSS void sample of \citet{cj15}.
The higher signal-to-noise void-galaxy cross-correlations (Fig.~\ref{fig:vg-log}) yield even more precise measurements of the void bias.
Void bias is constant over the range of scales we measure, outside twice the void radius.
Bias results (Fig.~\ref{fig:bias}) comparing auto- and cross-correlations are consistent in all size bins, spanning void radii over the range $15 - 30 \mpch$.
Using the cross-correlation, we find that void bias falls from $5.6 \pm 1.0$ to $-0.4 \pm 1.1$ over this range of radii, a $\sim 5\sigma$ detection of the trend.
In addition, we made a careful comparison with correlations in mock catalogs. We match the galaxy tracer density and bias and use the same void finder on the mocks. The void bias using both methods is consistent between data and mocks. Our results are promising for void cosmology with future surveys: based solely on the properties of tracer galaxies, mock catalogs provide predictions of large-scale void correlations that can be compared to measurements. 

With these mock voids, we choose different halo tracers to confirm previous results that void bias decreases with (i) increasing void size and (ii) increasing tracer density. The large, positive bias observed for small voids transitions through zero to increasingly negative values for large, densely sampled voids. 
Fig.~\ref{fig:vg-linear} shows a third trend: void samples defined from less biased halos show the same trend as with increasing tracer density. 
The trends with higher density or lower bias can be understood by noting that voids are defined at fixed underdensity threshold. So as the tracer density increases, we expect that the true mass density is lower for a region to be identified as a void. In summary:

High Density/Low bias of galaxy tracer\\
\-\hspace{0.7cm}$\rightarrow$ Emptier voids\\
\-\hspace{0.6cm} $\rightarrow$ Smaller amplitude positive bias for small voids\\
\-\hspace{0.7cm}$\rightarrow$ Higher amplitude negative bias for large voids

In future work, these trends can be tested and studied in more detail.
\citet{slh14} have shown that voids defined using denser tracer samples are more likely to accurately trace the dark matter field.
Similarly, it would be interesting to confirm whether or not our voids defined from less biased tracers are also more empty of dark matter \citep{nh15}.
In addition, higher order effects such as the detailed bias distribution of the void tracers may make a difference to the resulting void properties.
For example, as discussed in Sec.~\ref{sec:vv}, there is a slight $1-2\sigma$ discrepancy in the auto-correlation for some void size bins.
In future work we can explore whether using a mock sample with realistic mass-luminosity scatter and satellite fraction will remove any remaining differences between mock and data voids.
Another possible source of discrepancy is due to the different cosmologies: the \citet{planck15} best-fit value of $\Omega_{\rm m}$ is $\sim0.31 \pm 0.01$, whereas the simulations use 0.26.

We checked that differing treatmeants of redshift space distortions (RSD) in mocks and data is not the source of this discrepancy.
When analyzing the mock catalogs, the preceding results involve running the void finder and all measurements using real space coordinates.
However, we have separately analyzed the mocks after converting galaxy positions to redshift space using peculiar velocity information.
The correlation functions change by a negligible amount within our errors, as might be anticipated from \citet{lambas16}.
This work finds that void pairwise velocities are small, usually below 100 km/s, which would cause an $\sim 1\mpch$ shift along the x-axis of correlation function plots.
See also \citet{sef14} which finds that individual void macrocenters move very little over the course of the void's lifetime.

Galaxy samples in current surveys can be used for additional void clustering measurements. These include the higher density LOWZ and CMASS galaxy samples of BOSS discussed in Section 3.3. In the future, much larger surveys will be available. The majority of these surveys are imaging surveys that provide photometric redshifts, from which void samples can be created, but must be interpreted keeping in mind the uncertainty in the position along the line of sight. Three dimensional galaxy surveys from the Prime Focus Spectrograph on the Subaru telescope, and the Euclid and WFIRST satellites in the next decade will enable void studies over larger volumes and higher redshifts \citep{kcd13}.

A new approach to void cosmology involves the joint analysis of observables described in this paper and \citet{cj15}: void-galaxy, void-void, and void-mass correlations from weak lensing, as well as void abundances. These observables are analogous to cluster cosmology.
By running the same void finder on data and realistic void samples from simulations with varying cosmology, detailed comparisons between survey data and theoretical predictions can be made. The pros and cons of using voids as cosmological probes have been discussed in the literature. One obvious limitation is the lower statistical precision of void measurements. However there are theoretical reasons to expect a stronger signal for some cosmological tests, in particular modified gravity effects discussed below. For clustering measurements, 
more study is needed to understand the requirements for voids to provide useful tests.
Improvements to void finder algorithms and increasing amounts of data from ongoing surveys will only improve the signal-to-noise of these initial measurements of void-void, void-galaxy, and void-mass correlations.
A different cosmological test relies on the insensitivity of voids to nonlinear redshift space distortions. The varying bias levels studied here are relevant for redshift space distortions; in particular, the nearly unbiased void population may be useful in performing the Alcock-Pacynksi test \citep{ap79} to constrain cosmology \citep{hsl14,hsl15}.

Void lensing and clustering may be particularly useful for applications such as constraining modified gravity theories.
\citet{bcl15} have shown Galileon models have a large effect on void lensing predictions.
It would be interesting to test whether such theories also give significantly different results for void bias.
In addition, the changes to void number functions in chameleon theories \citep{ccl13,lam15,zivick15,cai15} may cause changes in the void bias as a function of void size.

\section*{Acknowledgments}
We are grateful to the referee for helpful comments which improved the final manuscript.
We would also like to thank Eric Baxter, Neal Dalal, Nico Hamaus, Mike Jarvis, Elisabeth Krause, Masahiro Takada and especially Ravi Sheth for helpful discussions.
Some of the results in this paper have been derived using the HEALPix \citep{ghb05} package.
JC and BJ are partially supported by the US Department of Energy grant de-sc0007901.
We benefitted from the hospitality of the Aspen Center for Physics, supported in part by National Science Foundation Grant No. PHYS-1066293.
CS was supported by the Spanish Ministerio de Economia y Competitividad (MINECO) under projects FPA2012-39684, and Centro de Excelencia Severo Ochoa SEV-2012-0234.



\end{document}